\documentclass[a4paper,12pt,UKenglish]{article}

\usepackage{amssymb,amsmath,natbib,amsthm}
\usepackage{graphicx,ifpdf,color} 
\usepackage[bottom=2.5cm, left=2cm, right=2cm, top=2cm]{geometry}
\usepackage[figuresright]{rotating}
\usepackage{subcaption}

\usepackage{bm}

\newcommand{\Sidak}{\v Sid\'ak }

\ifpdf 
   \DeclareGraphicsExtensions{.pdf,.eps,.png} 
\fi

\begin{document}

\title{Is the familywise error rate in genomics controlled by methods based on the effective number of independent tests?}

\author{K. K. Halle$^{1,2}$ S. Djurovic$^{3,4}$, O. A. Andreassen$^{5,6}$ and M. Langaas$^{1}$\\ \\
\footnotesize\parbox{.88\linewidth}{$^1$Department of Mathematical Sciences, Norwegian University of Science
and Technology, NO-7491 Norway.
$^2$Liaison Committee between the Central Norway Regional Health Authority (RHA) and the Norwegian University of Science and Technology (NTNU), Trondheim, Norway.
$^3$NORMENT, K. G. Jebsen Centre for Psychosis Research, Department of Clinical Science, University of Bergen, Bergen, Norway.
$^4$Department of Medical Genetics, Oslo University Hospital, Oslo, Norway.
$^5$NORMENT, K. G. Jebsen Centre for Psychosis Research, Division of Mental Health and Addiction, Oslo University Hospital and Institute of Clinical Medicine, University of Oslo, Oslo, Norway.
$^6$Institute of Clinical Medicine,  University of Oslo, NO-0318 Oslo, Norway.}
}
\date{\today}
\maketitle

\begin{abstract}
In genome-wide association (GWA) studies the goal is to detect association between one or more genetic markers and a given phenotype. The number 
of genetic markers in a GWA study can be in the order hundreds of thousands and therefore multiple testing methods are needed. This paper presents a set of popular
methods to be used to correct for multiple testing in GWA studies. All are based on the concept of estimating an effective number of independent tests.    
We compare these methods using simulated data and data from the TOP study, and show that the effective number of independent tests is not additive over blocks of independent genetic markers unless we 
assume a common value for the local significance level. We also show that the reviewed methods based on estimating the effective number of independent tests in general do not 
control the familywise error rate.

\end{abstract}


\setcounter{section}{0}
\section{Introduction}
In genome-wide association (GWA) studies, many genetic markers are tested for association with a given phenotype. The number of genetic markers
can be in the order hundreds of thousands and effective methods to adjust for multiple testing is a central topic when analyzing GWA data. Methods used to adjust for multiple testing need to take into
account that genetic markers in general are correlated and also be able to adjust for possible confounding factors such as population structure \citep{Price2006}. 
Single-step methods can control the familywise error rate (FWER) at level $\alpha$ by estimating a local significance level, $\alpha_{\text{loc}}$, which can be defined as the cut-off value 
for detecting significance. The Bonferroni method estimates the local significance level by $\alpha_{\text{loc}}=\alpha/m$, where $m$ is the number of tests and $\alpha$ 
is the chosen value of the familywise error rate. When the tests are dependent, the Bonferroni method is known to be conservative. The commonly used local significance level $\alpha_{\text{loc}}=5\cdot  10^{-8}$ for GWA studies \citep{Risch1996} is motivated by a Bonferroni correction based on one million independent genetic markers. The \Sidak method assumes the tests are independent and estimates 
the local significance level by $\alpha_{\text{loc}}=1-(1-\alpha)^{1/m}$. The maxT permutation method by \cite{WestfallYoung1993} gives strong control of the FWER when the subset pivotality condition is satisfied \citep{Meinshausen}. 
Permutation methods such as the maxT method need valid permutations. A valid permutation does not change the joint distribution of the test statistics corresponding to the true null hypotheses \citep{Goeman2014}. Permutation methods are computationally intensive for large datasets, such as GWA data.

Different methods for control of the overall error rate in GWA studies have been developed. A number of these methods are based on estimating an effective number of 
independent tests, $M_{\text{eff}}$,  which is subsequently used to find the local significance level. Methods for estimating the effective number of independent tests have been developed by \cite{cheverud}, \cite{nyholt},
\cite{LiJi2005}, \cite{gao1} and \cite{Galwey2009} among others. These methods estimate an effective number of independent tests, $M_{\text{eff}}$, from the genotype data and
then use the estimated $M_{\text{eff}}$ in either the Bonferroni or \Sidak method to estimate the local significance level. In this paper, we will
discuss the concept of estimating an effective number of independent tests in GWA studies, with focus on the relationship between the effective
number of independent tests, the local significance level and the overall FWER. We will compare the $M_{\text{eff}}$-based methods with the method presented by \cite{Halle2016}. We will use computational examples and data from a genome-wide association (GWA) study to illustrate and compare the different methods. 

This paper is organized as follows. In Section \ref{sec:background} we will present theory about multiple testing and matrix algebra. 
In Section \ref{sec:meffmethods} we will define the concept of an effective number of independent tests. In Section \ref{sec:meff} different methods for estimating the effective number of independent tests will be presented and the methods will be compared in Section \ref{sec:results}. The paper will conclude with a discussion in Section \ref{sec:discussion} and conclusion in Section 
\ref{sec:conclusion}.

\section{Statistical background}\label{sec:background}

In this section we will present notation and set-up for testing for genotype-phenotype association and some background theory on multiple testing correction. 

\subsection{Notation and data}
We assume that phenotype, $m$ genetic markers and $d$ environmental covariates are available from $n$ independent individuals. 
Let $\bm{Y}$ be an $n$-dimensional vector with the phenotype variable. Let $X_{\text{e}}$ be an $n \times d$ matrix of environmental covariates (with ones in the first column corresponding to an intercept), and $X_{\text{g}}$ 
an $n \times m$ matrix of genetic markers, then $X = (X_{\text{e}} X_{\text{g}})$ is an $n \times (d+m)$ covariate matrix. The genetic data are assumed to be from common variant biallelic genetic markers
with alleles $a$ and $A$, where $A$ is the minor allele based on the estimated minor allele frequency. We use additive coding $0,1,2$ for the three possible genotypes
$aa, Aa$ and $AA$, respectively. 

\subsection{Modelling genotype-phenotype associations using generalized linear models}\label{sec:glm}

We assume that the relationship between the genetic markers and the phenotype can be modelled using a generalized linear model (GLM) \citep{glm}, 
where the $n$-dimensional vector of linear predictors is
\begin{align*}
\bm{\eta}=X_{\text{e}}\bm{\beta}_e+X_{\text{g}}\bm{\beta}_g = X\bm{\beta} 
\end{align*}
where $\bm{\beta}=(\bm{\beta}_\text{e}^{\text{T}} \bm{\beta}_\text{g}^{\text{T}})^{\text{T}}$ is a $d+m$-dimensional unknown parameter vector. The link function $g$, defined by $\eta_i=g(\mu_i)$ is assumed to be canonical, implying that the contribution to the log likelihood for observation $i$ is $l_i=(Y_i\eta_i-b(\eta_i))/\phi_i+c(Y_i,\phi_i)$, where $b$ and $c$ are functions defining the exponential family of the phenotype, $Y_i$, and $\phi_i$ is the dispersion parameter. In our context, $\phi_i=\phi$ for all observations. For $Y_i$ normally distributed, $g(\mu_i)=\mu_i$ and $\phi=\sigma_i^2=\sigma^2$, and for $Y_i$ Bernoulli distributed, $g(\mu_i)=\log(\frac{\mu_i}{1-\mu_i})$ and $\phi=1$, where
$\mu_i=\text{E}(Y_i)$ and $\sigma_i^2=\text{Var}(Y_i)$.  

The score vector for testing the null hypothesis $H_0: \bm{\beta}_\text{g}=\bm{0}$ is given by
\begin{align}
\bm{U} = \frac{1}{\phi}X_{\text{g}}^T(\bm{Y}-\hat{\bm{\mu}}_\text{e}),
\label{scorevector}
\end{align}
where $\phi$ is the dispersion parameter and $\hat{\bm{\mu}}_\text{e}$ contains the fitted values from the null model with only the environmental covariates, $X_{\text{e}}$ (see e.g. \cite{Halle2016}). The vector $\bm{U}_{\text{g}|\text{e}}$ is asymptotically normally distributed with mean $\bm{0}$ and covariance matrix $V_{\text{g}|\text{e}}=\frac{1}{\phi^2}(X_{\text{g}}^T\Lambda X_{\text{g}}-X_{\text{g}}^T\Lambda X_{\text{e}}(X_{\text{e}}^T \Lambda X_{\text{e}})^{-1}X_{\text{e}}^T \Lambda X_{\text{g}})$ \citep{Halle2016}, where $\Lambda$ is a diagonal matrix with $\text{Var}(Y_i)$ on the diagonal.  

We are not interested in testing the complete null hypothesis $H_0: \bm{\beta}_\text{g}=\bm{0}$, instead we are interested in testing the null hypothesis
$H_{0j}: \beta_{\text{g}j}=0$ for each genetic marker $j, j=1,\ldots, m$. 
We consider the standardized components of the score vector, $\bm{T}=(T_1, \ldots, T_m)$, where
\begin{align}
T_j = \frac{\bm{U}_{\text{g}|\text{e} j}}{\sqrt{V_{\text{g}|\text{e} jj}}}, j = 1,\ldots, m. 
\label{Tk}
\end{align}
Note that the dispersion parameter $\phi$ is cancelled in the test statistics, but the elements of $\Lambda$ need to be estimated.
Each component $T_j, j=1,\ldots, m$ is asymptotically standard normally distributed and the vector $\bm{T}$ is asymptotically 
multivariate normally distributed $\bm{T} \sim N_m(\bm{0}, R)$, where the elements of the covariance matrix $R$ are 
$\text{Cov}(T_i,T_j)=\frac{V_{\text{g}|\text{e} \text{ } ij}}{\sqrt{V_{\text{g}|\text{e} \text{ } ii}V_{\text{g}|\text{e} \text{ } jj}}}$ where $V_{\text{g}|\text{e} \text{ } ii}$ is element $ii$ of the matrix $V_{\text{g}|\text{e}}$. 

As previously shown by us if the genetic and environmental covariates have near zero Pearson correlations, the correlations between score test statistics are equal to the genotype correlations between the genetic markers \citep{Halle2016},
\begin{align}
\text{Cor}(T_i,T_j) \approx \text{Cor}(X_{\text{g}i},X_{\text{g}j}), i=1,\ldots,m \text{ and } j=1,\ldots,m. 
\label{corrT}
\end{align}

The methods presented in Section \ref{sec:meff} use $\text{Cor}(X_{\text{g}i},X_{\text{g}j})$ and not $\text{Cor}(T_i,T_j)$ in the calculation of $\alpha_{\text{loc}}$.
This means that the results presented in this paper are valid for GLMs without environmental covariates (except intercept) and models where the genetic markers and
environmental covariates have zero Pearson correlation.

\subsection{Methods for control of the familywise error rate}

We consider a multiple testing problem where $m$ genetic markers are tested for association with a given phenotype. 
In GWA studies only a few significant results are expected and therefore we consider methods to control the familywise error rate (FWER). The FWER is defined as
\begin{align*}
\text{FWER}=P(V>0)
\end{align*}
where $V$ is the number of false positive results. Following the notation in \cite{Halle2016}, we denote by $O_j$ the event that the null hypothesis for genetic marker $j, j =1,\cdots, m$ is
not rejected and the complementary event is denoted $\bar{O}_j$. The event $O_j$ is of the form $|T_j| < d$ where $T_j$ is the score test statistic and $P(\bar{O}_j)=2\Phi(-d)=\alpha_{\text{loc}}$ is
the asymptotic probability of false rejection of the null hypothesis for genetic marker $j,j=1,\cdots,m$ where $\Phi$ is the univariate standard normal cumulative distribution function. 
Then, $P(O_j)=1-\alpha_{\text{loc}}$ where $\alpha_{\text{loc}}$ is the cut-off value used to detect significance. The FWER can be written as
\begin{align}
\text{FWER}=P(\bar{O}_1 \cup \cdots \cup \bar{O}_m) = 1-P(O_1 \cap \cdots \cap O_m)
\label{fwer}
\end{align}
under the complete null hypothesis. Note that the joint probability $P(O_1 \cap \cdots \cap O_m)$ depends on the value of the local significance level, $\alpha_{\text{loc}}$. 

The Bonferroni method is valid for all types of dependence structure between the test statistics and
the local significance level is found using the union formulation for the FWER as in Equation \eqref{fwer}. Using Boole's inequality
\begin{align*}
\alpha = \text{FWER}=P(\bar{O}_1 \cup \cdots \cup \bar{O}_m) \leq \sum_{j=1}^m P(\bar{O}_j) = \sum_{j=1}^m \alpha_{\text{loc}}=m\alpha_{\text{loc}}
\end{align*}
and the local significance level is $\alpha_{\text{loc}} = \frac{\alpha}{m}$ for the Bonferroni method. The Bonferroni method is known to be conservative when the test statistics are dependent. 
If the tests are assumed independent we find the \Sidak correction by solving 
Equation \eqref{fwer}
\begin{align*}
\text{FWER}=1-P(O_1 \cap \cdots \cap O_m) = 1-\prod_{j=1}^m P(O_j)
\end{align*}
which gives the local significance level for the \Sidak method
\begin{align}
\alpha_{\text{loc}} = 1-(1-\alpha)^{1/m}. 
\label{sidak}
\end{align}

\cite{Halle2016} presented an efficient and powerful alternative to the Bonferroni and \Sidak method. This method is computationally 
efficient compared to permutation methods, and can be used when the exchangeability assumption is not satisfied \citep{Halle2016}. Background theory about exchangeability and regression models is found in \cite{Commenges2003}. The method is based on the score test and approximating the joint probability $P(O_1 \cap \cdots \cap O_m)$ in Equation \eqref{fwer} by several integrals of low dimension. 
If the vector of test statistics follows a multivariate normal distribution with correlation matrix, $R$, $\bm{T}\sim N_m(\bm{0},R)$ and the number of genetic markers is less than or equal to $1000$, the numerical integration method described by \cite{Genz1992,Genz1993} can up to some level of accuracy be used to calculate the high dimensional integral in Equation \eqref{fwer}. This method is implemented in the R package mvtnorm \citep{mvtnorm}

\section{The effective number of independent tests}\label{sec:meffmethods}

In this section we will present the concept of an effective number of independent tests, $M_{\text{eff}}$, 
and show how $M_{\text{eff}}$ is used to correct for multiple testing in GWA studies. From Equation \eqref{sidak}, assuming $m$ independent tests and the FWER level $\alpha$, 
we find $\alpha_{\text{loc,m}} = 1-(1-\alpha)^{1/m}$.
The relationship between $m$, $\alpha_{\text{loc,m}}$ and $\alpha$ can be rewritten as 
\begin{align*}
m = \frac{\log(1-\alpha)}{\log(1-\alpha_{\text{loc,m}})}.
\end{align*}
For a general multiple testing problem where the FWER level $\alpha$ and the local significance level $\alpha_{\text{loc}}$ is known, the effective number of independent tests, $M_{\text{eff}}$, can be expressed as  \citep{MoskvinaSchmidt2008}
\begin{align}
M_{\text{eff}}=\frac{\log(1-\alpha)}{\log(1-\alpha_{\text{loc}})}. 
\label{meffdefinition}
\end{align}
If the $m$ tests are dependent, $M_{\text{eff}}<m$ and $\alpha_{\text{loc}}>\alpha_{\text{loc,m}}$, and if the tests are independent, $M_{\text{eff}}=m$ and $\alpha_{\text{loc}}=\alpha_{\text{loc,m}}$.
Note that the effective number of independent tests depends on both the familywise error rate, $\alpha$, and the local significance level, $\alpha_{\text{loc}}$. 
If the FWER can be expressed in terms of $\alpha_{\text{loc}}$, e.g. by use of Equation \eqref{fwer}, $\alpha_{\text{loc}}$ can be found by solving the equation $\text{FWER}=\alpha$. 

If $\alpha_{\text{loc}}$ and $\alpha$ are known, we can use Equation \eqref{meffdefinition} to calculate the effective number of 
independent tests. The methods presented in Section \ref{sec:meff} estimates $M_{\text{eff}}$ from the genotype correlation matrix and then provide a value of $\alpha$ to calculate $\alpha_{\text{loc}}$ using Equation \eqref{meffdefinition}.

\subsection{Independent blocks}\label{sec:additivity}
When the number of genetic markers is less than or equal to $1000$ and the vector of test statistics follows a multivariate normal distribution, numerical methods can be used to calculate the high dimensional integral that results from Equation \eqref{fwer} to some level of accuracy, for example by the method of \cite{Genz1992,Genz1993}. When analyzing GWA data, the number of genetic markers is much larger than $1000$. 
In this case, we can use an approximation of the high dimensional integral as done by \cite{Halle2016} or divide the data into independent blocks based on the genetic structure such that the high dimensional integral becomes a product of several lower dimensional integrals. 

Assume the genetic markers can be divided into $B$ independent blocks, that is, the events $\{O_1,\cdots,O_mÊ\}$ are divided into $B$ independent blocks, \\
$\{O_1,\cdots,O_{m_1}Ê\}, \{O_{m_1+1},\cdots,O_{m_2} \}, \cdots, \{O_{m_{B-1}+1},\cdots,O_mÊ\}$  so that $O_{i}$ and $O_{j}$ are independent if they belong to different blocks. 
Equation \eqref{fwer} can now be written as
\begin{align*}
\text{FWER} 	&= 1-P(O_1 \cap \cdots \cap O_m) \\
			&= 1-P(O_1 \cap \cdots \cap O_{m_1}Ê) \cdots P(O_{m_{B-1}+1} \cap \cdots \cap O_m),
\end{align*}
that is, the $m$-dimensional integral is a product of $B$ lower dimensional integrals. If we define the FWER for block $b$ as $\alpha_b = 1-P(O_{m_{b-1}+1} \cap \cdots \cap O_{m_b})$ we can write Equation \eqref{fwer} as 
\begin{align*}
\text{FWER} &= 1-\prod_{b=1}^B(1-\alpha_b).
\end{align*}

We are interested in a common local significance level, $\alpha_{\text{loc}}$, for all blocks. Using Equation \eqref{meffdefinition}, we can write
\begin{align}
M_{\text{eff}} 	&= \frac{\log(1-\alpha)}{\log(1-\alpha_{\text{loc}})} \nonumber \\
			&= \frac{\log P(O_1 \cap \cdots \cap O_m)}{\log(1-\alpha_{\text{loc}})} \nonumber  \\
			&= \frac{\log(P(O_1 \cap \cdots \cap O_{m_1}Ê)\cdots P(O_{m_{B-1}+1} \cap \cdots \cap O_m))}{\log(1-\alpha_{\text{loc}})} \nonumber \\
			&= \frac{\sum_{b=1}^B\log(1-\alpha_b)}{\log(1-\alpha_{\text{loc}})} 
\label{meffsum}
\end{align}  
If we define the effective number of independent tests for block $b$ as
\begin{align*}
M_{\text{eff},b} = \frac{\log(1-\alpha_b)}{\log(1-\alpha_{\text{loc}})}, b=1,\ldots, B
\end{align*}
we get $M_{\text{eff}}=\sum_{b=1}^B M_{\text{eff},b}$. 
Note that the common local significance level, $\alpha_{\text{loc}}$, needs to be calculated simultaneously for all $B$ blocks. 

Alternatively, \cite{Stange2016} used $\alpha_b=1-(1-\alpha)^{1/B}$ and denoted the effective number of independent tests for block $b,b=1,\cdots,B$ by
\begin{align*}
M_{\text{eff},b} = \frac{\log(1-\alpha_b)}{\log(1-\alpha_{\text{loc,b}})}, b=1,\ldots, B
\end{align*}
where $\alpha_{\text{loc,b}}$ is the local significance level for block $b$. 
With all $O_j$ of the form $|T_j|<c$, the $\alpha_b$ will depend on the dependency structure in the block and the number of markers in each block are typically different. However, this definition will give different local significance levels, $\alpha_{\text{loc,b}}$, for different blocks. 

\section{Methods for estimating the effective number of independent tests}\label{sec:meff}

The Bonferroni and \Sidak method can easily correct for multiple testing irrespective of the number of genetic markers. The Bonferroni method is known to be conservative when test statistics are dependent, and the \Sidak method may be invalid when the test statistics are dependent. Resampling methods can be used to approximate the high dimensional integral, but they are computationally intensive when the number of genetic markers is large. 

Another solution to correct for multiple testing in GWA studies was introduced by \cite{cheverud}, with the concept of an effective number of independent tests, $M_{\text{eff}}$. Various methods for calculating the effective number of independent tests have been suggested and in this section, the methods of \cite{cheverud}, \cite{nyholt}, \cite{gao1}, \cite{LiJi2005} and \cite{Galwey2009} are presented. These methods are all based on first estimating an effective number of independent tests and then use Equation \eqref{sidak} to calculate the local significance level, $\alpha_{\text{loc}}$.

\subsection{Eigenvalues of the genotype correlation matrix}\label{sec:eigenvalues}
Let $X_{\text{g}}^{*}$ be the $n \times m$ centred and scaled genotype matrix (see Appendix \ref{sec:matrixalgebra})
and assume no missing data for the genotypes. 

The singular value decomposition of the matrix $X_{\text{g}}^{*}$ is
\begin{align*}
X_{\text{g}}^{*}=U D V^T
\end{align*}
where $U$ is a $n \times r$ matrix, $D$ is a $r \times r$ matrix and $V$ is a $m \times r$ matrix \citep[Appendix A]{Mardia1979}. The columns of the matrix $U$
are the $n$ eigenvectors of the matrix $X_{\text{g}}^{*}X_{\text{g}}^{*T}$. The columns of the matrix $V$ are the $m$ eigenvectors of the matrix $X_{\text{g}}^{*T}X_{\text{g}}^{*}$.
The matrices $X_{\text{g}}^{*T}X_{\text{g}}^{*}$ and $X_{\text{g}}^{*}X_{\text{g}}^{*T}$ have the same nonzero eigenvalues but different eigenvectors. $D$ is a diagonal matrix with the singular values of $X_{\text{g}}^{*}$ as diagonal elements. 
The estimated genotype correlation matrix is the $m \times m$ matrix
\begin{align*}
\hat{R}=\frac{1}{n-1}X_{\text{g}}^{*T}X_{\text{g}}^{*}
\end{align*}
The maximal number of nonzero eigenvalues of the correlation matrix is equal to $\text{rank}(X_{\text{g}}^{*})=r \leq \min(n-1,m)$.

\subsection{Methods for estimating $M_{\text{eff}}$}

The methods presented in this section are all based on the eigenvalues of the genotype correlation matrix, $\hat{R}$, and are not related to the statistical test used. 

The methods of \cite{cheverud} and \cite{nyholt} use a simple interpolation between the two extreme cases of complete independence ($M_{\text{eff}}=m$) and complete
dependence ($M_{\text{eff}}=1$) betweeen the genetic markers. 
In the first case, $\hat{R}$ is the identity matrix, and all eigenvalues are $1$, so that the variance of the eigenvalues is $0$. In the second case, all entries of $\hat{R}$ have absolute value $1$, and every column is a multiple of any column. Thus, the rank of $\hat{R}$ is $m-1$, so that $0$ is an eigenvalue of multiplicity $m-1$. In addition, $m$ is a simple eigenvalue. The variance of the eigenvalues is $m$. Interpolation yields \citep{cheverud}
\begin{align*}
M_{\text{eff}}=m\left(1-(m-1)\frac{\text{Var}(\lambda)}{m^2} \right)
\end{align*}
where $\text{Var}(\lambda)$ is the empirical variance of the eigenvalues of the matrix $\hat{R}$. This was by \cite{nyholt} rewritten as 
\begin{align*}
M_{\text{eff}}=1+(m-1)\left(1-\frac{\text{Var}(\lambda)}{m} \right)
\end{align*}
where $\text{Var}(\lambda)$ is the empirical variance of the eigenvalues of the matrix $\hat{R}$.  
The local significance level for the method of Cheverud and Nyholt is found by first calculating $M_{\text{eff}}$ and then replacing $m$ with $M_{\text{eff}}$ in the \Sidak correction. 
Cheverud used genotype data, while Nyholt used the correlation matrix based on haplotype data 
and also removed all genetic markers in perfect  linkage disequilibrium except one before estimating the effective number of independent tests \citep{nyholt2005}. 

The method of \cite{gao1} uses the composite linkage disequilibrium (CLD) to calculate the pairwise correlation matrix between the genetic markers. The genotype correlation between the genetic markers is an estimate of two times the CLD \citep{Halle2012}. The method of Gao uses the eigenvalues of the matrix $\hat{R}$ to estimate $M_{\text{eff}}$. The eigenvalues of $\hat{R}$ are sorted in decreasing order, $\lambda_1 \geq \lambda_2 \geq \ldots \geq \lambda_m$. 
and $M_{\text{eff}}$ is the number of eigenvalues to explain a given percent of the variation in the data, that is,
\begin{align*}
\frac{1}{m}\sum_{i=1}^{M_{\text{eff}}-1} \lambda_i < c < 	\frac{1}{m}\sum_{i=1}^{M_{\text{eff}}} \lambda_i					
\end{align*}
where $c$ is a predetermined cut-off. \cite{gao1} used $c=0.995$, which means that $M_{\text{eff}}$ is the number of eigenvalues which explain $99.5\%$ of the variation in the data. 
The method of \cite{gao1} consists in finding the $M_{\text{eff}}$ and then $m$ is replaced with $M_{\text{eff}}$ in the Bonferroni method to find the local significance level. Later, \Sidak correction was used instead \citep{gao2}. When the number of markers, $m$, is larger than the sample size $n$, the
genetic markers are divided into blocks of smaller size.  \cite{Halle2012} showed that the method of Gao is highly dependent on the block size used.  The method is also dependent on the chosen value of $c$,
and there is no given connection between $c$ and the FWER level $\alpha$. 

Other methods for estimating $M_{\text{eff}}$ based on the matrix $\hat{R}$ include the methods of  \cite{LiJi2005} and \cite{Galwey2009}. 
The method of Li and Ji estimate effective number of independent tests by
\begin{align*}
M_{\text{eff}}=\sum_{i=1}^m f(|\lambda_i|), 
\end{align*}
where $f(x)=I(x\geq 1)+(x-\lfloor x \rfloor ), x\geq 0$ and $I(x\geq 1)$ is the indicator function. 
The method of \cite{Galwey2009} is described as an improvement of the method of \cite{LiJi2005} and the estimate the effective number of independent tests is defined as
\begin{align*}
M_{\text{eff}}=\frac{(\sum_{i=1}^m \sqrt{\lambda_i})^2}{\sum_{i=1}^m \lambda_i}. 
\end{align*}
The method of Galwey can also be used when the matrix $\hat{R}$ is not positive semidefinite. When we have missing data for 
some of the observations, $\hat{R}$ is estimated based on pairwise complete observations and may not be positive semidefinite. 
The method of Galwey assumes that all negative eigenvalues are small in absolute value and therefore are set to zero. 

The methods presented in the section are all based on first estimating the effective number of independent tests, $M_{\text{eff}}$, and then choose a value $\alpha$ for the FWER to calculate the value of the local significance level, $\alpha_{\text{loc}}$, by Equation \eqref{meffdefinition}. In a previous paper, we have presented an alternative method to correct for multiple testing in GWA studies \citep{Halle2016}. This method, the Order $k$ method, estimates the local significance level, $\alpha_{\text{loc}}$, by approximating the high dimensional integral in Equation \eqref{fwer} by several integrals of lower dimension. Then, Equation \eqref{meffdefinition} can be used to calculate $M_{\text{eff}}$.

\section{Results}\label{sec:results}
In this section, we present results for estimating the effective number of independent tests and the local significance level using correlation matrices with compound symmetry correlation structure, autoregressive order 1 (AR1) correlation structure and tridiagonal structure, see Appendix \ref{sec:computationalexamples}. These correlation structures all contain a correlation parameter, $\rho$. We also consider real data from the TOP study \citep{TOP1,TOP2}. For the different methods discussed in this paper, we compare the estimated local significance level and the estimated FWER between methods. 

\subsection{Computational examples}

We assume the vector of test statistics follows a multivariate normal distribution with correlation matrix, $R$, $\bm{T}\sim N_m(\bm{0},R)$. When the correlation matrix, $R$, is of compound symmetry structure, the high dimensional integral in Equation \eqref{fwer} can be written as a 
product of univariate integrals \cite[p.~17]{genzbretzbok},
\begin{align}
P(O_1 \cap \cdots \cap O_m) = \int_{\mathbb{R}}\phi(y)\prod_{i=1}^m\left[\Phi\left(\frac{d-\sqrt{\rho}y}{\sqrt{1-\rho^2}}\right)-\Phi\left(\frac{-d-\sqrt{\rho}y}{\sqrt{1-\rho^2}}\right)\right]dy
\label{fwercs}
\end{align}
where $d=\Phi(1-\frac{\alpha_{\text{loc}}}{2})$.  For AR1 correlation matrices, the $m$-dimensional integral in Equation \eqref{fwer} can only be simplified
to a $m-1$-dimensional integral. Also, for tridiagonal matrices, to our knowledge, there does not exist a simple form of the $m$-dimensional integral and therefore, for AR1 and tridiagonal matrices, we compare our results to results obtained by using numerical integration using the R package mvtnorm \citep{mvtnorm} and the GenzBretz algorithm of \cite{Genz1992,Genz1993}. 
The method of Genz and Bretz can up to some level of accuracy be used to calculate the high dimensional integral in Equation \eqref{fwer} for arbitrary correlation matrices when the number of genetic markers is  $m \leq 1000$. We use this algorithm with absolute error tolerance $abseps=10^{-9}$ in all examples. 
For each of the methods presented in Section \ref{sec:meff}, and the Order $k$ method (with $k=2$) presented by \cite{Halle2016}, we found the estimated effective number of independent tests and estimated local significance level based on the FWER level $\alpha=0.05$. The method of \cite{Halle2016} controls the FWER at level $\alpha$ when the vector of test statistics is asymptotically multivariate normally distributed with a correlation matrix of compound symmetry or AR1 structure \citep{PhDthesis}. The value of the parameter, $\rho$, of the three special correlation matrices were chosen such that the correlation matrices are positive definite, that is, for tridiagonal matrices we consider $\rho \leq 0.5$. For compound symmetry correlation matrices, we also considered $\rho>0.4$ to be unrealistic for large blocks of genetic markers. 

\subsubsection{The local significance level}

We are interested in solving Equation \eqref{fwer} with $\text{FWER}=\alpha$ to estimate a value of the local signficance level, $\alpha_{\text{loc}}$. 
We consider examples where the vector of test statistics follows a multivariate normal distribution with correlation matrix, $R$, $\bm{T}\sim N_m(\bm{0},R)$ and $m=1000$. 
When $R$ is of compound symmetry structure we use the formula in Equation \eqref{fwercs} to calculate the high dimensional integral in Equation \eqref{fwer} and when
$R$ is of AR1 or a tridiagonal structure we use the numerical integration method of \cite{Genz1992,Genz1993} with high precision. 
We will denote the local significance level found using either Equation \eqref{fwercs} or by numerical integration as $"$True$"$ in the examples in this section. 
 
Table \ref{tab:simalphaloc} show the estimated local significance level for different correlation matrices with $m=1000$ genetic markers using the methods presented in this paper. The relationship between the local significance level, $\alpha_{\text{loc}}$, and the effective number of independent tests, $M_{\text{eff}}$, is given in Equation \eqref{meffdefinition}. 

\begin{table}[h!]
\begin{center}
\vspace{0.5cm}
\begin{tabular}{lccccccc}
		&				&		&				&				&					&				&		\\
			&	$\rho$ 		& True	& Cheverud  		& Galwey			& Li and Ji 			& Gao		 	& Order 2 	\\ 
\hline
C.S.		 			& $0.1$			& $5.3367$ & $5.1810$ 		& $5.5914$		& $5.6928$			& $5.1550$		& $5.1297$		\\ 
					& $0.2$			& $6.0797$ & $5.3427$ 		& $6.2253$		& $6.4035$ 			& $5.1602$ 		& $5.1318$   		\\
					& $0.3$			& $7.6416$ & $5.6359$ 		& $7.0466$ 		& $7.3169$ 			& $5.1654$ 		& $5.1387$ 		\\
					& $0.4$			& $10.5939$ & $6.1050$ 		& $8.1392$ 		& $8.5343$ 			& $5.1706$ 		& $5.1575$ 		\\
\hline
AR1					& $0.1$    			& $5.1288$ & $5.1293$ 		& $5.1550$ 		& $5.1292$ 			& $5.1602$ 		& $5.1297$        	\\
					& $0.2$    			& $5.1313$ & $5.1296$ 		& $5.2351$ 		& $5.1292$ 			& $5.1654$ 		& $5.1318$       	\\
					& $0.3$    			& $5.1370$ & $5.1302$ 		& $5.3781$ 		& $5.1292$ 			& $5.1758$ 		& $5.1387$       	\\
					& $0.4$   			& $5.1647$ & $5.1312$ 		& $5.6012$ 		& $5.8552$ 			& $5.1862$ 		& $5.1575$        	\\
					& $0.5$   			& $5.2080$ & $5.1326$ 		& $5.9371$ 		& $6.1134$ 			& $5.2020$ 		& $5.2034$         	\\
					& $0.6$    			& $5.3286$ & $5.1350$ 		& $6.4498$ 		& $6.9407$ 			& $5.2285$ 		& $5.3085$         	\\
					& $0.7$   			& $5.5776$ & $5.1391$ 		& $7.2813$ 		& $8.2594$ 			& $5.2769$ 		& $5.5439$        	\\
					& $0.8$   			& $6.2465$ & $5.1474$ 		& $8.8235$ 		& $10.097$ 			& $5.3653$ 		& $6.0986$        	\\
					& $0.9$   			& $8.7265$ & $5.1731$ 		& $12.787$ 		& $14.489$ 			& $5.6614$ 		& $7.7456$        	\\
\hline
Trid.					& $0.1$			& $5.1289$ & $5.1293$  		& $5.1552$		& $5.1292$			& $5.1602$		& $5.1297$		\\
					& $0.2$			& $5.1308$ & $5.1296$		& $5.2375$		& $5.1292$			& $5.1706$		& $5.1318$		\\
					& $0.3$			& $5.1374$ & $5.1301$		& $5.3928$		& $5.1292$			& $5.1915$		& $5.1387$		\\
					& $0.4$			& $5.1722$ & $5.1308$		& $5.6675$		& $5.1292$			& $5.2553$		& $5.1575$		\\
					& $0.5$			& $5.2305$ & $5.1318$		& $6.3251$		& $5.1292$			& $5.9920$		& $5.2034$				\\
\end{tabular}
\caption{$10^{5}$ times estimated local significance level, $\alpha_{\text{loc}}$, for different correlation matrices (C.S. = compound symmetry, AR1=autoregressive order 1, Trid.=tridiagonal).}
\label{tab:simalphaloc}
\end{center}
\end{table}

\begin{figure}[h!]
    \centering
    \includegraphics[width=0.8\textwidth]{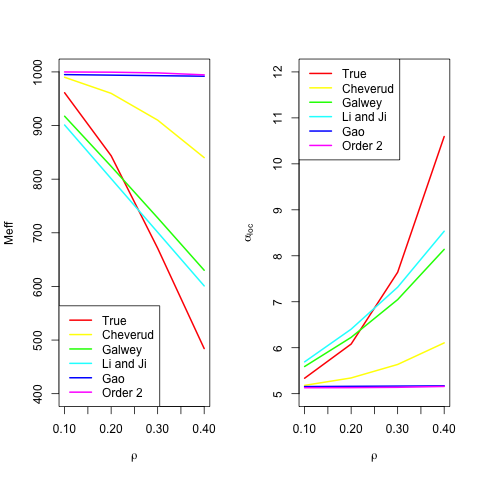}
    \caption{The estimated local significance level and effective number of independent tests for compound symmetry correlation matrices.}
    \label{fig:simcs}
\end{figure}

 Figure \ref{fig:simcs} shows the estimated local significance level and effective number of independent tests for correlation matrices with compound symmetry structure and $m=1000$ genetic markers. The methods which have lines crossing the line for the true value in Figure \ref{fig:simcs} will control the FWER only for some values of $\rho$. 

\begin{figure}[ht!]
    \centering
    \includegraphics[width=0.8\textwidth]{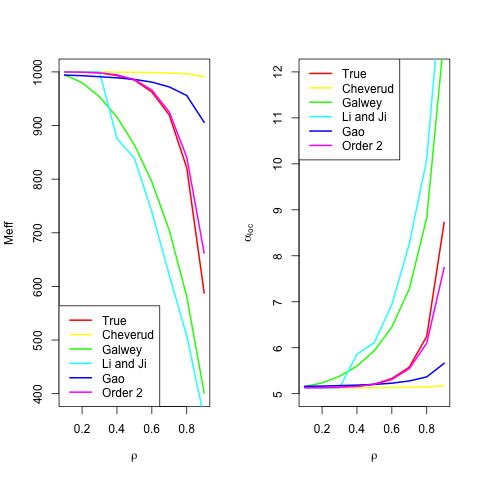}
    \caption{The estimated local significance level and effective number of independent tests for AR1 correlation matrices.}
    \label{fig:simdataar1matrix}
\end{figure}

Figure \ref{fig:simdataar1matrix} shows the estimated local significance level and effective number of independent tests for correlation matrices with AR1 structure  and $m=1000$ genetic markers. The correlation parameter was chosen as $\rho \in [0.1-0.9]$ and from Figure \ref{fig:simdataar1matrix} we see that the Order 2 method gives results closest to the true value. The methods of \cite{LiJi2005} and \cite{Galwey2009} do not control the FWER at level $\alpha=0.05$.

\begin{figure}[ht!]
    \centering
    \includegraphics[width=0.8\textwidth]{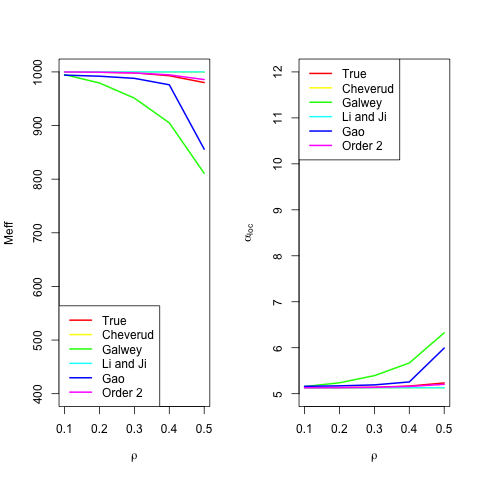}
    \caption{The estimated local significance level and effective number of independent tests for tridiagonal matrices with constant correlation $\rho$  on the first off-diagonal.} 
    \label{fig:simtridiagonal1}
\end{figure}

Figure \ref{fig:simtridiagonal1} shows the estimated local significance level and the effective number of independent tests for a tridiagonal band matrix with correlation $\rho$ on the first off-diagonal and $m=1000$ genetic markers. From Figure \ref{fig:simtridiagonal1} we see that the Order 2 method gives results closest to the true value. The methods of \cite{LiJi2005} and \cite{Galwey2009} do not control the FWER at level $\alpha=0.05$ for all values of $\rho$. 

\subsubsection{Estimated FWER}

In this section we compare the different methods by the estimated FWER. The estimated FWER is calculated using the numerical integration algorithm by \cite{Genz1992,Genz1993} given the estimated value of the local significance level from Table \ref{tab:simalphaloc}. We used absolute error tolerance $abseps=10^{-9}$ for the GenzBretz algorithm. 
 
Table \ref{tab:simfwer} shows the estimated FWER for different correlation matrices. 
For compound symmetry, the methods by \cite{cheverud}, \cite{gao1} and the Order 2 method of \cite{Halle2016} are conservative for all values of $\rho$, while the methods of \cite{Galwey2009} and \cite{LiJi2005} do not control the FWER at level $\alpha=0.05$ for all values of $\rho$. 
For autoregressive order 1 (AR1) correlation matrices the method of \cite{cheverud} is conservative for all values of $\rho$, the methods of \cite{Galwey2009} and \cite{LiJi2005} does not control the FWER at level $\alpha$ and that the method of \cite{gao1} is conservative for large values of $\rho$. The Order 2 method of \cite{Halle2016} controls the FWER at level $\alpha=0.05$ for all values of $\rho$, but is conservative for large values of $\rho$. 
For tridiagonal correlation matrices we see that the methods of \cite{Galwey2009}  and \cite{gao1} do not control the FWER at level $\alpha=0.05$. The other methods control the FWER at level $\alpha=0.05$.

\begin{table}[ht!]
\begin{center}
\vspace{0.5cm}
\begin{tabular}{lcccccr}
Correlation 	&				&				&				&					&				&		\\
structure		&	$\rho$ 		& Cheverud  		& Galwey			& Li and Ji 			& Gao		 	& Order 2 	\\ 	
\hline
Compound	 &  $0.1$		& $0.0487$ 	& $0.0522$  & $0.0531$ & $0.0484$ & $0.0482$  \\ 
symmetry 		&  $0.2$		& $0.0447$ 	& $0.0512$  & $0.0525$ & $0.0431$ & $0.0429$  \\ 
&  $0.3$		& $0.0388$ 	& $0.0467$  & $0.0485$ & $0.0355$ & $0.0355$    \\  
&  $0.4$		& $0.0321$	& $0.0403$  & $0.0419$  & $0.0281$ & $0.0278$   \\ 
 \hline
AR1 &   $0.1$ 		& $0.0500$  &$0.0502$  &$0.0500$ &$0.0503$ &$0.0500$ \\Ê
&  $0.2$ 		& $0.0500$  &$0.0510$  &$0.0500$ &$0.0503$ &$0.0500$ \\Ê
&  $0.3$ 		& $0.0499$  &$0.0523$  &$0.0499$ &$0.0504$ &$0.0500$ \\Ê
&  $0.4$ 		& $0.0497$  &$0.0542$  &$0.0565$ &$0.0503$ &$0.0500$ \\
&  $0.5$ 		& $0.0493$  &$0.0567$  &$0.0584$ &$0.0499$ &$0.0499$ \\
&  $0.6$ 		& $0.0483$  &$0.0602$  &$0.0645$ &$0.0492$ &$0.0500$ \\
&  $0.7$ 		& $0.0460$  &$0.0642$  &$0.0724$ &$0.0472$ &$0.0495$ \\
&  $0.8$ 		& $0.0415$  &$0.0686$  &$0.0779$ &$0.0433$ &$0.0485$ \\Ê
&  $0.9$ 		& $0.0309$  &$0.0715$  &$0.0800$ &$0.0338$ &$0.0455$ \\  
\hline
Tridiagonal &  $0.1$		& $0.0500$  & $0.0502$  & $0.0500$  & $0.0503$	& $0.0500$ \\
&  $0.2$		& $0.0500$  & $0.0510$  & $0.0500$  & $0.0504$	& $0.0500$ \\
&  $0.3$		& $0.0499$  & $0.0524$  & $0.0499$  & $0.0505$	& $0.0500$ \\
&  $0.4$		& $0.0498$  & $0.0548$  & $0.0498$ & $0.0509$	& $0.0500$ \\
\end{tabular}
\caption{Estimated FWER for different correlation matrices}
\label{tab:simfwer}
\end{center}
\end{table}

\subsection{Independent blocks}\label{sec:independentblocks}

The effective number of independent tests for independent blocks was discussed in Section \ref{sec:additivity}. We have shown that the effective number of independent tests is additive over independent blocks of genetic markers when assuming a common value of $\alpha_{\text{loc}}$. 

To illustrate this we consider an example with $m=1000$ genetic markers, divided into 100 blocks, each of 10 genetic markers with a compound symmetry correlation structure with $\rho=0.7$ in each block. 
We want to control the FWER at level $\alpha=0.05$. First, we estimate the effective number of independent tests for each block, and then sum these estimates to get a total effective number of independent tests, $\sum_{b=1}^B M_{\text{eff,b}}$. These results are shown 
in Table \ref{tab:simblocks}. The $M_{\text{eff}}$ is then transformed into $\alpha_{\text{loc}}$ and we calculated the FWER for the different methods using the R package mvtnorm \citep{mvtnorm} with the numerical integration method by \cite{Genz1992,Genz1993}. The methods of \cite{cheverud}, \cite{Galwey2009} and \cite{LiJi2005} do not control the FWER at level $\alpha=0.05$. The Order 2 method of \cite{Halle2016} is developed to be used separately for independent blocks. 

\begin{table}[h!]
\begin{center}
\vspace{0.5cm}
\begin{tabular}{lcc}
		&         	$\sum_{b=1}^B	M_{\text{eff},b}$	& FWER	 \\
\hline
Cheverud 		&    $559.00$  	& 0.0687  \\
Galwey		&    $582.38$  	& 0.0667  \\
Li and Ji		&    $400.00$  	& 0.0926  \\
Gao			&    $1000.00$  & 0.0407 \\
\end{tabular}
\caption{Estimated $M_{\text{eff}}$ and FWER calculated from block-wise estimates.}
\label{tab:simblocks}
\end{center}
\end{table}

It is also possible to use the methods on the full correlation matrix. 
Table \ref{tab:simblocks1} shows the effective number of independent tests and the FWER, calculated using the correlation matrix for all $m$ genetic markers. Also, in this case, the methods of Galwey and Li and Ji do not control the FWER. The method of \cite{cheverud} does not control the FWER when using the sum of the block-wise estimates (Table \ref{tab:simblocks}), but the method is conservative when we use all genetic markers to estimate the effective number of independent tests (Table \ref{tab:simblocks1}). The Order 2 method gives results closest to the FWER level $\alpha=0.05$. 

\begin{table}[h!]
\begin{center}
\vspace{0.5cm}
\begin{tabular}{lcccc}
		&          $M_{\text{eff}}$	& FWER \\
\hline
Cheverud 		&     $995.59$   & 0.0401 \\
Galwey		&     $582.38$   & 0.0658 \\
Li and Ji		&     $400.00$   & 0.0933 \\
Gao			&     $984.00$   & 0.0401 \\
Order 2		&     $932.72$   & 0.0430 \\
\end{tabular}
\caption{Estimated $M_{\text{eff}}$ and FWER using the full correlation matrix.}
\label{tab:simblocks1}
\end{center}
\end{table}

\subsection{The TOP study}\label{sec:topdata} 

We studied GWA data from the TOP study \citep{TOP1,TOP2}. This data contains genetic information for 672972 genetic markers for 1148 cases (with schizophrenia and bipolar disorder) and 420 controls. To illustrate the different methods discussed in this paper, we consider one block of size $m=1000$ consisting of the first 1000 genetic markers (based on position on the chip used for genotyping) from chromosome 22. 
Table \ref{tab:topresults} shows the estimated effective number of independent tests and corresponding local significance level for 1000 genetic markers on chromosome 22 in the TOP data. As for the previous examples in this section, we compare our results with the numerical integration method of \cite{Genz1992,Genz1993} which gives the estimated local significance level $\alpha_{\text{loc}}=7.3049\cdot 10^{-5}$ and a corresponding effective number of independent tests $M_{\text{eff}}=702.15$. Table \ref{tab:topresults} shows the results using other methods for estimating the local significance level or the effective number of independent tests. Using the numerical integration method of \cite{Genz1992,Genz1993}, we find the corresponding FWER level for each of these methods using the correlation matrix from the TOP data. 

\begin{table}[h!]
\begin{center}
\vspace{0.5cm}
\begin{tabular}{lccr}
Method	&	$M_{\text{eff}}$		& $ 10^{5}\alpha_{\text{loc}}$ 				& FWER 		\\
\hline
Cheverud		&	$993.5$		&	$5.1626$		& $0.0351$	\\
Order 2		& 	$985.7$		& 	$5.2034$		& $0.0358$	\\
Gao			&	$615.0$		&	$8.3400$		& $0.0567$	\\
Li and Ji		& 	$424.0$		& 	$12.0967$	& $0.0776$	\\
Galwey		&	$396.5$		&	$12.9343$	& $0.0830$ 	\\
\end{tabular}
\caption{Estimated local significance level and FWER using 1000 genetic markers from the TOP data, chromosome 22.}
\label{tab:topresults}
\end{center}
\end{table}

\section{Discussion}\label{sec:discussion}
We have discussed the concept of using an estimated number of independent tests, $M_{\text{eff}}$, to correct for multiple testing in GWA studies. 
We have shown that $M_{\text{eff}}$ depends on both the local significance level and the FWER, and that the effective number of independent tests is additive over independent blocks of genetic markers only 
when assuming a common value of $\alpha_{\text{loc}}$ for each block. Different methods for estimating $M_{\text{eff}}$ were presented in Section \ref{sec:meff} and compared using 
computational examples and real data in Section \ref{sec:results}. 

The methods of \cite{cheverud}, \cite{nyholt}, \cite{gao1}, \cite{LiJi2005} and \cite{Galwey2009} are all based on estimating $M_{\text{eff}}$, using the eigenvalues of the genotype correlation matrix and then replacing $m$ with  
$M_{\text{eff}}$ in the \Sidak correction to find the local significance level, $\alpha_{\text{loc}}$. These methods are  
not related to the statistical test used, and can therefore not include adjustment for confounding factors such as for example
population structure in GWA studies. These methods will also give the same value of $M_{\text{eff}}$ for all values of the FWER level $\alpha$
since the algorithms used to calculate $M_{\text{eff}}$ include $\alpha$ only when calculating $\alpha_{\text{loc}}$.

The method of \cite{gao1} is based on the eigenvalues of the genotype correlation matrix. When the number of genetic markers is larger than the sample size, $n$, the rank of the genotype correlation matrix
is at most $n-1$ (see Section \ref{sec:eigenvalues}), which means the maximal number of nonzero eigenvalues is $n-1$. The maximal effective number of independent tests using this method in this case is $n-1$. 
When the number of genetic markers is large, the genotype correlation matrix is divided into independent blocks of smaller size, but as discussed by \cite{Halle2012}, the results of this method is also highly dependent on the block size used. The method of Gao also depends on the parameter $c$ which is used to find $M_{\text{eff}}$. 

 \cite{cheverud} and \cite{gao1} find the total $M_{\text{eff}}$ by the sum of $M_{\text{eff}}$ estimates for smaller, independent blocks and the local significance level for the whole dataset is found using the \Sidak correction with the total estimate of the effective number of independent tests. As discussed in Section \ref{sec:additivity} and Section
\ref{sec:independentblocks}, it is not possible to find block-wise estimates of $M_{\text{eff}}$ which sums to the total $M_{\text{eff}}$, without assuming a common and known  
value of the local significance level. Table \ref{tab:simblocks} shows that the method of \cite{cheverud} is not additive as the sum of the block-wise estimates is
$M_{\text{eff}}=559.0$, while the estimated number of independent tests using the whole genotype correlation matrix is $M_{\text{eff}}=995.6$. 

\section{Conclusion}\label{sec:conclusion}
In this paper we have presented and discussed the concept of using an effective number of independent tests, $M_{\text{eff}}$, to correct for multiple testing in GWA studies. 
We have seen that $M_{\text{eff}}$ depends on both the local significance level and the FWER, and that $M_{\text{eff}}$ is additive over independent blocks of genetic markers only 
when assuming a common value of the local significance level, $\alpha_{\text{loc}}$ for the blocks. 
Different methods were compared using computational examples with different correlation structures as well as real data from the TOP study and we have seen that the Order 2 method presented by \cite{Halle2016} controls
the FWER in all examples. The other methods considered in this paper (except the Bonferroni and \Sidak methods) fail to control the FWER in at least one of the examples studied.

\section*{Software}
The statistical analysis were performed using the statistical software R \citep{R}. R code for the examples in this paper are available at http://www.math.ntnu.no/$\sim$karikriz.

\section*{Acknowledgements}
The PhD position of the first author is founded by the Liaison Committee between the Central Norway Regional Health Authority (RHA) and the 
Norwegian University of Science and Technology (NTNU). \\
\emph{Conflict of Interest:} None declared.

\bibliographystyle{chicago}
\bibliography{meff}

\appendix

\section{Matrix algebra}\label{sec:matrixalgebra}

In this section we will present some background theory about singular value decomposition and eigenvalues, which give a theoretical 
background for the methods presented in Section \ref{sec:meff}. 
We let $X_{\text{g}}^{*}$ be the $n \times m$ centered (and scaled) genotype matrix
and assumes no missing data for the genotypes. 
The elements of the matrix $X_{\text{g}}^{*}$ are
\begin{align*}
X_{g,ij}^{*} = \frac{X_{g,ij} -\bar{X}_{g,.j}}{(n-1)\sqrt{\frac{1}{n-1}\sum_{k=1}^n(X_{g,ij}-\bar{X}_{g,.j})^2}}, i = 1,\ldots, m,j = 1,\ldots, n,
\end{align*}
where $\bar{X}_{g,.j}=\frac{1}{n}\sum_{k=1}^n X_{g,kj}$ is the mean value for genetic marker $j$.

The estimated genotype correlation matrix is the $m \times m$ matrix
\begin{align*}
\hat{R}=\frac{1}{n-1}X_{\text{g}}^{*T}X_{\text{g}}^{*}
\end{align*}
with elements ($i = 1,\ldots, m,j = 1,\ldots, n$)
\begin{align*}
\hat{R}_{ij} &= \sum_{k=1}^n\frac{1}{n-1}\big(\frac{X_{g,ki}^{*}-\bar{X}_{g,.i}^{*}}{\sqrt{\frac{1}{n-1}\sum_{k=1}^n(X_{g,ki}^{*}-\bar{X}_{g,.i}^{*})^2}}\big)\big( \frac{X_{g,kj}^{*}-\bar{X}_{g,.j}^{*}}{\sqrt{\frac{1}{n-1}\sum_{k=1}^n(X_{g,kj}^{*}-\bar{X}_{g,.j}^{*})^2}} \big).
\end{align*}
The diagonal elements of $\hat{R}$ are
\begin{align*}
\hat{R}_{ii}	&= \frac{1}{n-1}\sum_{k=1}^n \big(\frac{X_{g,ki}^{*}-\bar{X}_{g,.i}^{*}}{\sqrt{\frac{1}{n-1}\sum_{k=1}^n(X_{g,ki}^{*}-\bar{X}_{g,.i}^{*})^2}}\big)^2 \\
			&= \frac{1}{n-1}\sum_{k=1}^n \frac{(X_{g,ki}^{*}-\bar{X}_{g,.i}^{*})^2}{\frac{1}{n-1}\sum_{k=1}^n(X_{g,ki}^{*}-\bar{X}_{g,.i}^{*})^2} \\
			&= \frac{\sum_{k=1}^n (X_{g,ki}^{*}-\bar{X}_{g,.i}^{*})^2}{\sum_{k=1}^n(X_{g,ki}^{*}-\bar{X}_{g,.i}^{*})^2} \\
			&= 1, i=1,\ldots, m.
\end{align*}
We denote the $r$ nonzero eigenvalues of $\hat{R}$ by $d_1, \ldots, d_r$. The sum of the $r$ nonzero eigenvalues is
\begin{align*}
\sum_{i=1}^r d_i = \sum_{i=1}^n d_i = \text{trace}(\hat{\rho})=m
\end{align*}
if $m < n-1$.

\section{Computational examples}\label{sec:computationalexamples}

In this section we present some additional results for the computational examples in Section \ref{sec:results}. The FWER level is $\alpha=0.05$ in all examples.

\subsection{Compound symmetry correlation matrix}
A compound symmetry correlation matrix with $m=5$ and parameter $\rho$ is given in \eqref{cs}.
\begin{align}
R = \begin{bmatrix}
       1 & \rho & \rho	& \rho	& \rho        \\[0.3em]
       \rho & 1 & \rho 	& \rho	& \rho 	\\[0.3em]
       \rho & \rho & 1	& \rho	& \rho	\\[0.3em]
       \rho & \rho & \rho & 1		& \rho	\\[0.3em]
       \rho & \rho & \rho & \rho 	& 1 
     \end{bmatrix}
     \label{cs}
\end{align}

For a compound symmetry correlation matrix with correlation coefficient $\rho$, the $m$ eigenvalues are $\lambda_1=1+(m-1)\rho$ and $\lambda_2,\cdots, \lambda_m=1-\rho$. 
As noted in Section \ref{sec:meff}, \cite{cheverud} estimated the effective number of independent tests by 
\begin{align*}
M_{\text{eff}}=m\left(1-(m-1)\frac{\text{Var}(\lambda)}{m^2} \right)
\end{align*}
For a compound symmetry correlation matrix with $m=1000$ and $\rho=0.1$, we have $\lambda_1=100.9$ and $\lambda_2,\cdots, \lambda_{1000}=0.9$. 
The sample variance of these eigenvalues is $\text{Var}(\lambda)=10$, which using the method of \cite{cheverud} gives
\begin{align*}
M_{\text{eff}} 	&= m\big(1-(m-1)\frac{\text{Var}(\lambda)}{m^2} \big) \\
			&= 1000\big(1-(1000-1)\frac{10}{1000^2} \big) \\
			&= 990.01
\end{align*}
and with FWER level $\alpha=0.05$ we can calculate the local significance level as shown in Table \ref{tab:simalphaloc}. 
For the method of \cite{gao1} the first eigenvalue is $\lambda_1=100.9$ and 
\begin{align*} 
\lambda_1/\sum_{i=1}^{1000} \lambda_i=100.9/1000=0.10,
\end{align*}
that is, the first eigenvalue only explains $10\%$ of the variance in the data. 
We need $995$ eigenvectors to explain $99.5\%$ of the variation in the data, so the effective number of independent tests estimated by the method of \cite{gao1} is $M_{\text{eff}}=995$. 

The method of \cite{LiJi2005} estimate the effective number of independent tests by
\begin{align*}
M_{\text{eff}}=\sum_{i=1}^m f(|\lambda_i|), 
\end{align*}
where $f(x)=I(x\geq 1)+(x-\lfloor x \rfloor ), x\geq 0$ and $I(x\geq 1)$ is the indicator function. For a compound symmetry correlation matrix with $\rho=0.1$ and $m=1000$ genetic markers, this gives
\begin{align*}
M_{\text{eff}} 	&= \sum_{i=1}^m f(|\lambda_i|) \\
			&= 1+\sum_{i=1}^m (\lambda_i-\lfloor \lambda_i \rfloor ) \\
			&= 1+\sum_{i=1}^m (1-\rho-\lfloor 1-\rho \rfloor ) \\
			&= 1+\sum_{i=1}^m (1-\rho) \\
			&= (1+m)-m\rho \\
			&= 1+1000-1000\cdot 0.1 \\
			&= 901.
\end{align*}
\cite{Galwey2009} estimate the effective number of independent tests by 
\begin{align*}
M_{\text{eff}}=\frac{(\sum_{i=1}^m \sqrt{\lambda_i})^2}{\sum_{i=1}^m \lambda_i}. 
\end{align*}
For a compound symmetry correlation matrix with $\rho=0.1$ and $m=1000$ genetic markers, this gives
\begin{align*}
M_{\text{eff}}	&= \frac{(\sum_{i=1}^m \sqrt{\lambda_i})^2}{\sum_{i=1}^m \lambda_i} \\
			&= \frac{(\sqrt{1+(m-1)\rho}+\sum_{i=2}^m \sqrt{1-\rho})^2}{m} \\
			&= \frac{(\sqrt{1+(m-1)0.1}+\sum_{i=2}^m \sqrt{1-0.1})^2}{1000} \\
			&= 917.34. 
\end{align*}

\subsection{Autoregressive order 1 (AR1) correlation matrix}
An AR1 correlation matrix with $m=5$ and parameter $\rho$ is given in \eqref{ar1}.
\begin{align}
R = \begin{bmatrix}
       1 	& \rho & \rho^2	& \rho^3		& \rho^4        \\[0.3em]
       \rho 	& 1 & \rho 	& \rho^2		& \rho^3 	\\[0.3em]
       \rho^2	& \rho & 1	& \rho		& \rho^2	\\[0.3em]
       \rho^3 & \rho^2 	& \rho & 1		& \rho	\\[0.3em]
       \rho^4 & \rho^3 	& \rho^2 & \rho 	& 1 
     \end{bmatrix}
     \label{ar1}
\end{align}

\subsection{Tridiagonal band matrices with constant correlation}
A tridiagonal correlation matrix with $m=5$ and parameter $\rho$ is given in \eqref{tridiag1}.
\begin{align}
R = \begin{bmatrix}
       1 	& \rho 	& 0		& 0		& 0	        \\[0.3em]
       \rho 	& 1 		& \rho 	& 0		& 0	 	\\[0.3em]
       0	 & \rho	 & 1		& \rho	& 0		\\[0.3em]
       0	 & 0		 & \rho 	& 1		& \rho	\\[0.3em]
       0	 & 0		 & 0		 & \rho 	& 1 
     \end{bmatrix}
     \label{tridiag1}
\end{align}

\end{document}